\documentclass[aps, prb, superscriptaddress, reprint, showpacs, showkeys, preprintnumbers, amsmath, amssymb, longbibliography]{revtex4-1}
\usepackage{graphicx}
\usepackage{float}

\begin{document}

\title{Three-particle electron-hole complexes in two-dimensional electron systems}

\author{V.~A. Kuznetsov}
\email{To whom correspondence may be addressed. volod\_kuzn@issp.ac.ru}
\affiliation{National Research University Higher School of Economics, 141700 Moscow, Russia}
\affiliation{Institute of Solid State Physics, Russian Academy of Sciences, 142432 Chernogolovka, Russia}
\author{L.~V. Kulik}
\affiliation{Institute of Solid State Physics, Russian Academy of Sciences, 142432 Chernogolovka, Russia}
\author{M.~D. Velikanov}
\affiliation{Moscow Institute of Physics and Technology, 141700 Dolgoprudniy, Russia}
\affiliation{Skolkovo Institute of Science and Technology, Nobel str. 3, 143026 Moscow, Russia}
\author{A.~S.~Zhuravlev}
\affiliation{Institute of Solid State Physics, Russian Academy of Sciences, 142432 Chernogolovka, Russia}
\author{A.~V. Gorbunov}
\affiliation{Institute of Solid State Physics, Russian Academy of Sciences, 142432 Chernogolovka, Russia}
\author{S.~Schmult}
\altaffiliation[Current address: ]{TU Dresden, Institute of Semiconductors and Microsystems, N{\"o}thnitzer Stra{\ss}e 64, 01187 Dresden, Germany}
\affiliation{Max-Planck-Institut f{\"u}r Festk{\"o}rperforschung, Heisenbergstra{\ss}e 1, 70569 Stuttgart, Germany.}
\author{I.~V. Kukushkin}
\affiliation{Institute of Solid State Physics, Russian Academy of Sciences, 142432 Chernogolovka, Russia}
\affiliation{National Research University Higher School of Economics, 141700 Moscow, Russia}


\date{\today}

\begin{abstract}
Three-particle complexes consisting of two holes in the completely filled zero electron Landau level and an excited electron in the unoccupied first Landau level are investigated in a quantum Hall insulator. The distinctive features of these three-particle complexes are an electron-hole mass symmetry and the small energy gap of the quantum Hall insulator itself. Theoretical calculations of the trion energy spectrum in a quantizing magnetic field predict that, besides the ground state, trions feature a hierarchy of excited bound states. In agreement with the theoretical simulations, we observe new photoluminescence lines related to the excited trion states. A relatively small energy gap allows the binding of three-particle complexes with magnetoplasma oscillations and formation of plasmarons. The plasmaron properties are investigated experimentally.
\end{abstract}

\maketitle

\section{Introduction}
Trions are coupled three-particle electron-hole states. Their existence in bulk semiconductors was predicted as far back as 1958 by Lampert \cite{Lampert1958}. The plasmaron, which is a particular trion case, can be represented as a bound state of a two-particle plasmon and a hole or an electron \cite{Hedin1967}. Here, the ``plasma'' component of the plasmaron energy can exceed the trion binding energy. In literature, there are reports on experimental observations of three-particle trions and plasmarons \cite{Shay1971,Tediosi2007}. The first recognized observation of a plasmaron in a two-dimensional (2D) system (graphene near the Dirac point) using angle-resolved photoemission spectroscopy (ARPES) was reported recently \cite{Bostwick2010}.

Later on, the scientific community paid much attention to the magnetic-field dependencies of optical absorption, reflection, and photoluminescence (PL) spectra of undoped 2D semiconductor heterostructures with quantum wells (QWs). Extra luminescence lines appearing under the Mott transition conditions were attributed to the trion consisting of a photoexcited exciton and a residual electron that enters the quantum well from the donor impurities in the heterostructure barrier \cite{Finkelstein1995,Shields1995}. However, according to results of theoretical study \cite{Dzyubenko00a} devoted to three-particle states in 2D systems, in a trion single-photon optical transitions are symmetry-forbidden. This introduces doubts regarding the soundness of the interpretation of the experimental results. The issue was addressed by performing a series of experimental studies on QWs with intentional donor doping at various distances from the quantum well center. The lines of the so-called trions were shown to be the lines of localized four-particle $D_{0X}$ complexes. These complexes are composed of a charged impurity in the quantum well barrier with an electron in QW bound to it ($D_0$ complex) and a photoexcited exciton bound to a $D_0$ complex \cite{Solovyev2009}. The proponents of the trion origin of extra lines based their counter-arguments on the fact that impurities in heterostructure barriers should be arranged randomly at an arbitrary distance from the quantum well. Therefore, a wide photoluminescence band that corresponds to all possible positions of barrier-ionized impurities should be observed, instead of a narrow line detected in the experiment. This argument was refuted in a different study, which did not explore four-particle $D_{0X}$ complexes, but investigated their forerunners instead, i.e., $D_0$ complexes, using the inelastic light-scattering technique. It appeared that in all of the quantum wells being considered, the unintentional impurities were not randomly arranged, but were placed in strict locations (in most cases, on the quantum well interface), which can be explained by the impurity-diffusion processes during heterostructure growth \cite{Zhuravlev2010}. Thus, the issue of observing translation-invariant (free) trions in the optical spectra of 2D systems was taken off.

The next step in the investigation of three-particle complexes was made in our recent works, where we developed new methods of forming dense magnetoexciton ensembles in 2D electron systems placed in a quantizing magnetic field $B$ \cite{Kulik2015,Kulik2016}. Under the conditions of Hall insulator with integer filling of Landau level, $\nu =2$, magnetoexcitons are two-particle electron-hole states composed of a Fermi hole (electron vacancy in a conduction band) in the filled zero electron Landau level and an excited electron in the empty first Landau level. There are two types of magnetoexcitons: a spin singlet with a zero total spin, and a spin triplet with a spin that is equal to one. In accordance with the Kohn theorem, the energy of the spin-singlet magnetoexciton with a zero momentum is equal to the cyclotron gap \cite{Kohn1961}. The spin-triplet magnetoexciton energy appears to be less than the cyclotron energy by the value of the Coulomb interaction energy, which is determined by the electron correlations \cite{Kulik2005}.

In contrast to the spin-singlet exciton, the spin-triplet magnetoexciton is not optically active (``dark'' magnetoexciton). However, a nonequilibrium ensemble of dark triplet magnetoexcitons can be created by the optical excitation of the electrons from the valence band to high-energy states of conduction band \cite{Kulik2015}. The main channel of changing the spin of the electron system is the valence-band heavy hole (Hh) spin-flip processes owing to the strong spin-orbit interaction in the GaAs valence band. Being excited to a higher Landau level ($>1$), the hole relaxes to the zero Hh-Landau level changing repeatedly its spin during relaxation. As the photoexcited Hh is further transformed into a Fermi hole by recombination of the conduction band electron with the photoexcited valence-band hole, the electron system changes its spin. Since the relaxation of spin-triplet magnetoexcitons to the ground state via the cyclotron photon emission, which is accompanied by the simultaneous change of the orbit and spin quantum numbers, is forbidden, the triplet magnetoexciton lifetime becomes extremely long. The complementary factor that extends the triplet magnetoexciton lifetime is that the energy minimum in the magnetoexciton dispersion is not at the zero momentum, $q = 0$, but at the momentum of the order of $1/l_B$ \cite{Kallin1984} (here $l_B = \sqrt{\frac{\hbar c}{e B}}$ is the magnetic length). Then, it becomes challenging to violate both the momentum and the spin conservation in all possible scattering processes that involve the relaxation of spin-triplet magnetoexcitons to the ground state. As a consequence, the relaxation time reaches very large values, up to 1~ms \cite{Kulik2016}. Considering that spin magnetoexciton lifetime is $10^7$ times larger than photoexcited hole recombination time (in the system in question the recombination times are about 100~ps), photoexcitation could enable us to achieve a high density of nonequilibrium spin-triplet magnetoexcitons $N_{X}$ (of the order of $10^{10}$ cm$^{-2}$).
\begin{figure}[h]
	\includegraphics{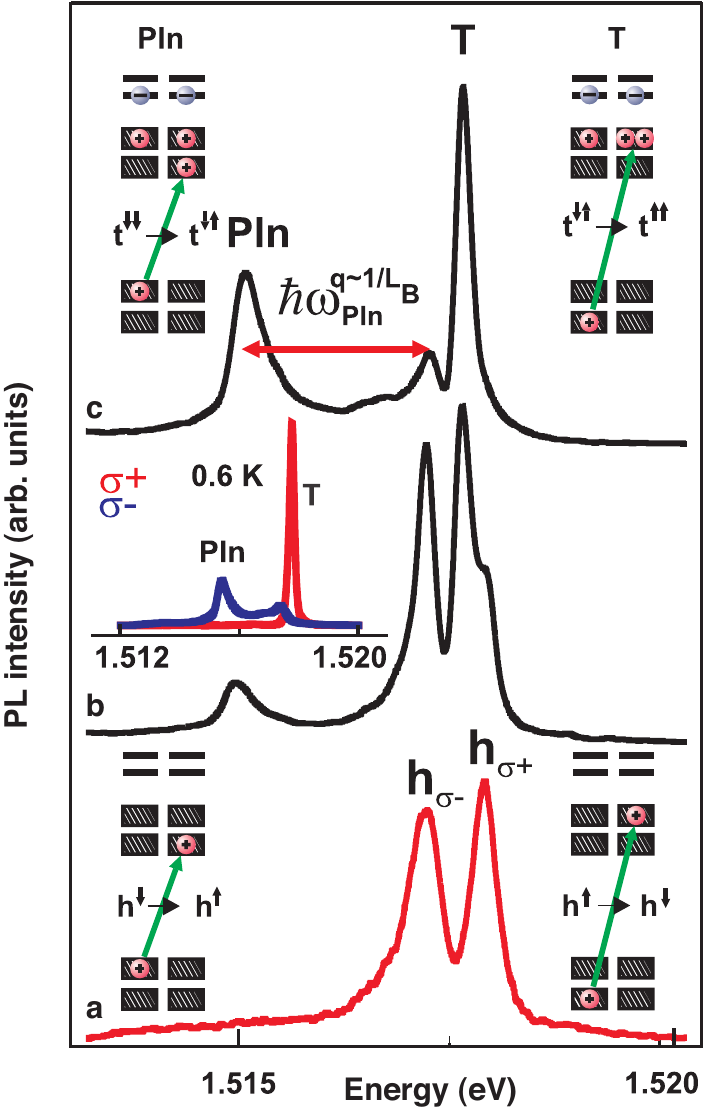}
	\caption{\label{fig:fig1}
		Schemes of single-particle and intra-trion recombination transitions in two emission polarizations, $\sigma^{-}$ and $\sigma^{+}$, and examples of PL spectra of the sample with electron concentration $n_{e}=2\cdot 10^{11}$ cm$^{-2}$ in the presence of a nonequilibrium ensemble of long-lived magnetoexcitons. (a) Bath temperature $T=$ 1.5~K (practically no triplet magnetoexcitons in the electron system). (b) $T=$ 0.6~K; the concentration of triplet magnetoexcitons $N_{X} < 0.01N_{\phi}$, the number of states in one electron spin Landau sublevel. (c) $T=$ 0.6~K; $N_{X}\approx0.1N_{\phi}$; here the lines of three-particle complexes, plasmarons (Pln) and trions (T), are dominant in the spectrum. The inset shows polarized PL spectra for the case (c).}
\end{figure}

In the presence of a nonequilibrium triplet magnetoexciton ensemble, the trion and plasmaron physics is unified within a single physical object, namely, the coupled state of  Fermi hole and magnetoexciton \cite{Zhuravlev2016}. Moreover, there are no symmetry restrictions on the observation of free (translation-invariant) three-particle complexes in photoluminescence spectra. A single photoexcited hole in the valence band is coupled to a magnetoexciton in the conduction band. Then, within the single-photon recombination process, it passes from the valence band to the Fermi hole state in the zero electron Landau level of the conduction band. The process can proceed with no change in the internal degrees of freedom of the three-particle complex. The final state of the recombination process is either a plasmaron in $\sigma^{-}$ polarization or a trion in $\sigma^{+}$ polarization, as shown in the diagrams of Figure~\ref{fig:fig1}. The difference in the properties of trions in two polarizations is owing to the fact that in $\sigma^{-}$ polarization, the spin quantum number of one of the Fermi holes coincides with that of the excited electron. The latter can occupy the Fermi hole state, transferring its excess energy to another electron-Fermi-hole pair, which is nothing but magnetoplasma oscillations in the presence of an extra Fermi hole. It should be noted that in contrast to the zero magnetic field case, where translations of the exciton, trion, and plasmaron centers of mass are described by the same quantum number (momentum), and where dependence of kinetic energy on momentum is continuous, in the presence of a magnetic field the situation is fundamentally different. The energy of magnetoexcitons that are neutral particles also has a continuous dependence on the generalized momentum. In contrast, the spectra of plasmaron and trion as charged quasiparticles are discrete.

The key point of three-particle states composed of conduction-band electrons and Fermi holes is that electron and hole masses are the same (mass symmetry). This is a rare phenomenon because electron and hole masses in semiconductors are usually different, and trion physics is treated similar to the physics of positive helium and negative hydrogen ions. The proposed system is much closer to positronium ion physics. This opens up a unique opportunity for direct experimental investigations of the novel hierarchy of bound energy states in a semiconductor object that is equivalent to a positive positronium ion in quantizing magnetic fields whose motion in one spatial direction is restricted by the external potential \cite{Xie2003}. Such an object is hardly realizable in standard conditions owing to a lack of positronium ions and to their short lifetimes \cite{Mills1981,Mills1983}. However, such objects can play a significant role in cosmological models \cite{Wheeler1946}.

\begin{figure}[h!]
	\includegraphics{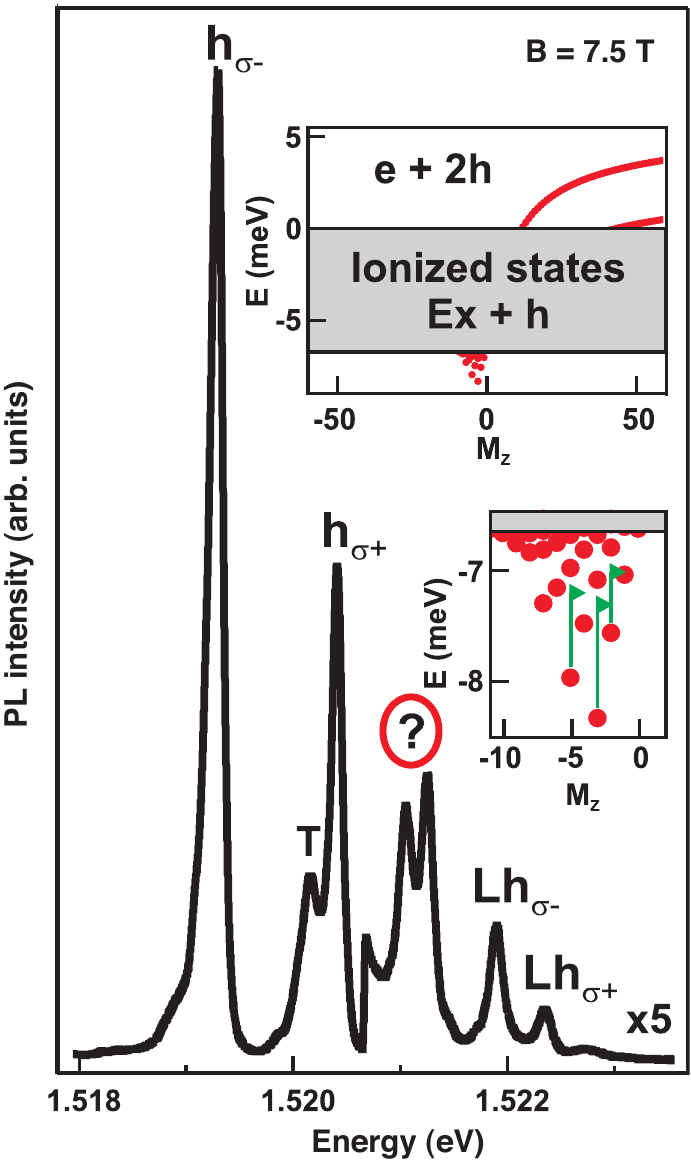}
	\caption{\label{fig:fig2}
		PL spectrum of the sample with $n_{e}=3.6\cdot 10^{11}$ cm$^{-2}$ in the presence of a nonequilibrium ensemble of long-lived magnetoexcitons. Besides the line of single-particle transitions from the valence band of heavy holes to the conduction band, the spectrum reveals an intra-trion transition line and a doublet of unknown lines. The upper inset shows a theoretical spectrum of three-particle states in $\sigma^{+}$ polarization calculated for the sample in question. The lower inset presents a magnified spectrum of three-particle states in the region of strongly coupled trion states. There is a small number of strongly coupled trion states with a binding energy around 1.5 meV. Weakly bound excited trion states pass to a continuum of unbound magnetoexciton and Fermi hole states. The inset also demonstrates possible transitions with $\Delta M_z = 0$.}
\end{figure}

Because excited electrons (first Landau level) and Fermi holes (zero Landau level) are described by a p-type and an s-type wave function, respectively, electron-hole symmetry is not complete. It is interesting to note that the intra Landau level energy spectrum for three-particle complexes with incomplete electron-hole symmetry is not exhausted by the ground state.  Under certain conditions, the spectrum is a hierarchy of bound excited states classified by a set of inner quantum numbers of trions. This hierarchy has nothing in common with any of the known spectral series, and the energy dependence of bound states does not follow any simple functional law \cite{Dzyubenko2000_ssc}. This is particularly remarkable considering that a trion consisting of three symmetrical s-electrons and holes (complete electron-hole symmetry) has exactly one bound ground state (spin-triplet in paired particles forming a trion) \cite{Palacios1996, Whittaker1997}. Neither bound excited states nor a plasmaron-type state occur in this case. In the same way, if the s-type wave function of one of the paired trion particles is replaced by a p-type wave function, and the unpaired particle is left with an s-type wave function, the trion will have exactly one bound ground state (again, spin-triplet in paired particles) \cite{Dzyubenko2000_ssc}. However, if the wave function of the unpaired trion particle is changed from the s-type to the p-type, and if both wave functions of paired particles are s-type, as is in our experimental case, then theoretical calculations yield a hierarchy of bound excited states. Most of them have a low binding energy, which makes it practically impossible to distinguish bound states from states of a continuum consisting of a magnetoexciton and a single Fermi hole. However, there is a small number of experimentally observable strongly bound excited states of trions (Figure~\ref{fig:fig2}).

In the work presented here, we continue experimental research into 2D three-particle complexes in a quantized magnetic field, and we report the observation of new lines in the photoluminescence spectra of a 2D electron system in the presence of a magnetoexciton ensemble. Such lines are observed simultaneously with trion lines in structures with high electron mobility. The lines exhibit energy exceeding that of the main optical transition in the trion, and lie in the energy region where there are no single-particle optical transistions. We provide evidence to support the fact that the new trion lines enter the range of optical transition from the excited trion states. Using original experimental techniques, we consider a plasmaron excitation spectrum, and show that there exists not a single plasmaron excitation, but an entire band of three-particle plasma excitations whose quantum numbers depend on the initial generalized momentum of the magnetoexciton forming the plasmaron. Therefore, one of the most unexpected results of the present work is the observation of the generalized momentum transfer from a neutral two-particle complex to a charged three-particle complex in the quantizing magnetic field, as well as evidence of the fact that there exists a band of plasmaron excitations whose energy values depend on the value of the transferred generalized momentum.

\section{Experiment}

High quality heterostructures with symmetrically doped GaAs/AlGaAs single quantum well were used to create nonequilibrium triplet magnetoexcitons and study PL spectra in the presence of a dense magnetoexciton ensemble. The electron concentration in the 2D channel $n_{e}$ did not exceed $3.6\cdot 10^{11}$ cm$^{-2}$, the dark mobility $\mu_{e}>8\cdot 10^6$ cm$^2/$V$\cdot$s. The QW widths varied within 30-35~nm. Narrower QWs ($20$ - $25$ nm) did not ensure electron mobility required for studies of three-particle complexes, and wider QWs (40~nm) did not provide required disequilibrium (relaxation time of nonequilibrium magnetoexcitons falls with increasing QW width owing to the flattening of the Coulomb minimum in the dispersion curve~\cite{Kallin1984}). The QW width optimum required for creating a nonequilibrium magnetoexciton ensemble at $\nu =2$ is reached within $25\div35$~nm. The relaxation time of nonequilibrium magnetoexcitons at other integer filling factors appears to be shorter. So, for now it turns out to be impossible to create quasi-stationary nonequilibrium systems with magnetoexciton concentration $N_{X}\sim10^{10}$ cm$^{-2}$ at integer filling factors other than $\nu=2$.

The samples were placed inside a pumped insert with liquid $^3$He which in turn was put into a $^4$He cryostat with a superconducting solenoid. Optical measurements were made at bath temperature $T$ down to 0.45~K using the double light fiber technique. One fiber was used for photoexcitation, the other for collecting emission from the sample and passing the PL signal onto the slit of a grating spectrometer equipped with a cooled CCD camera. A tunable TiSp-laser was used as an optical source for excitation of the electron system and creating a magnetoexciton ensemble. To measure spectra with specified circular polarization, $\lambda/4$ plate and linear polarizer retaining their polarization properties at liquid-helium temperature were placed between the sample and collecting light fiber. The sign of circular polarization was changed on the opposite one by reversing the direction of magnetic field.

The study of plasmarons was based on measurements of spatially resolved PL spectra. Samples were placed in liquid $^3$He inside a pumped insert equipped with an optical window for light input/output which in turn was put in a $^4$He-cryostat with a superconducting solenoid. The excitation source of the electron system was a tunable laser, its beam split into two: pump and probe beams. Using a lens system inside the $^3$He-insert, the laser beams were focused on the sample surface into two spatially separated round spots of 20 $\mu$m diameter each. The distance between the pump and probe spot centers was 200 $\mu$m. The pump beam power varied within two orders of magnitude: from 2 $\mu$W to 200 $\mu$W, while the probe beam power was kept constant and equal to 3 $\mu$W.

\section{Trions}

Next, we consider optical transitions of photoexcited holes from the valence band to the conduction band of the electron system in quantum well GaAs/AlGaAs in the presence of triplet magnetoexcitons. In a quantum Hall insulator at $\nu = 2$, a photoexcited hole may occur in magnetoexciton-free space. Such a situation is easily realizable at temperatures above 1.5~K when practically all magnetoexcitons are ``ionized'' (relaxed to the ground state by the resonant conversion of the magnetoexciton into a magnetoplasmon and recombination of the latter) \cite{Zhuravlev2016}. In this case, the recombination process involving the photoexcited hole becomes essentially single-particle recombination. The photoexcited hole in the valence band of the zero Hh-Landau level transforms into a single Fermi hole in the zero electron Landau level of the conduction band (Figure~\ref{fig:fig1}).

In the opposite case of a dense magnetoexciton ensemble with exciton density of the order of 0.1 of the electron states in the single-spin Landau level, $N_{X}\approx0.1N_{\phi}$, there remain practically no magnetoexciton-free sites. Therefore, the lines of three-particle complexes, trions and plasmarons, become dominant in the recombination spectra (Figure~\ref{fig:fig1}). The occurrence of major three-particle lines in the PL spectrum is accompanied by weaker extra lines (Figure~\ref{fig:fig2}) with optical transition energy values that are higher than the single-particle transition energy of photoexcited heavy holes in the zero Hh-Landau level. In addition, they are lower than the single-particle transition energy of photoexcited light holes (Lh) in the zero Lh-Landau level. The intensity of the new PL lines is appreciably higher than that of the lines of single-particle transitions from the zero Lh-Landau level. Thus, the new lines are related to transitions from the states occurring in the forbidden gap for single-particle electron-hole transitions.

\begin{figure}[H]
	\includegraphics{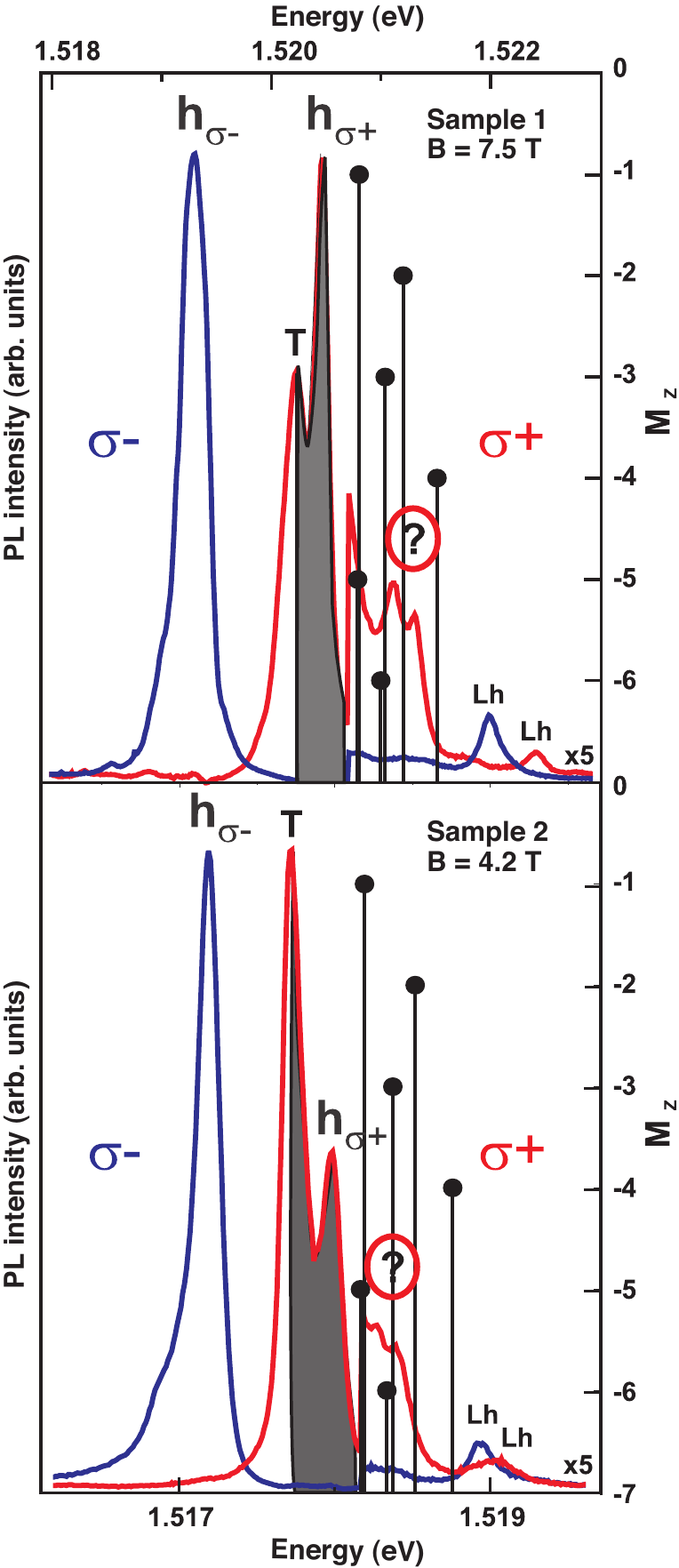}
	\caption{\label{fig:fig3}
		PL spectra measured in two polarizations, $\sigma^{-}$ (blue) and $\sigma^{+}$ (red), for the samples with concentration $n_{e}=3.6\cdot 10^{11}$ cm$^{-2}$ (top) and $n_{e}=2\cdot 10^{11}$ cm$^{-2}$ (bottom). Besides the lines of single-particle transitions from the valence bands of Hh and Lh, both cases reveal an intra-trion transition line (T) and lines of an unknown spectral doublet. The figure also shows the calculated spectrum of strongly bound trion states with $\Delta M_z=0$ plotted from the line of the main intra-trion transition (black dots). For each transition, the projection of the total momentum $M_z$ of the initial state is shown (right axis). The dark crosshatch indicates the energy range occupied by weakly bound trion states.}
\end{figure}

Then, the polarization properties of the new lines were investigated. The lines are observed in $\sigma^{+}$ polarization, where trion optical transitions should be observed (Figure~\ref{fig:fig3}); however, their energy is significantly greater than that of the major optical transition for trions. To determine whether the new lines fall into the region of the feasible optical transitions from the excited trion states, the trion energy spectrum was calculated considering the real potential profile limiting electron motion in the QW growth direction.

The problem of a charged-particle system in a magnetic field $B$ was considered in the work\cite{Avron1978}. The existence of a motion integral analogous to the total momentum for the case of $B = 0$ was established. The motion of a neutral two-particle system was separated on the center-of-mass motion and the internal motion around the center of mass \cite{Lamb1952, Gorkov1967}. Although in the general case of three interacting particles, the complete separation of internal and external degrees of freedom is impossible, their partial separation became possible in the case of equal cyclotron masses of all the particles~\cite{Dzyubenko2000_ssc}. We used the method proposed by A. Dzyubenko to calculate the trion excited states. It would be more convenient to work in the basis generated by relevant creation operators rather than in the coordinate representation. We consider the single-particle Hamiltonian for a 2D charged particle in an external magnetic field $\mathbf{B} = (0, 0, B)$, which is perpendicular to a 2D system. 

\begin{equation}\label{ham}
\mathcal{H} = \frac{\mathbf{\hat{\Pi}}^2}{2 m},
\end{equation}
where $\mathbf{\hat{\Pi}} = -i \hbar \nabla - e \mathbf{A}(\mathbf{r})$ is the particle momentum. In the cylindrical calibration, $\mathbf{A} = \frac{1}{2} \mathbf{B} \times \mathbf{r}$. We take creation and annihilation operators as well as related basis wave functions in the form

\begin{equation}
A^\dagger_e(\mathbf{r}) = \frac{1}{\sqrt{2}} \left(  \frac{z}{2 l_B} - 2 l_B \frac{\partial}{\partial z ^{*}}\right),
\end{equation}
\begin{equation}
B^\dagger_e(\mathbf{r}) = \frac{1}{\sqrt{2}} \left(  \frac{z^*}{2 l_B} - 2 l_B \frac{\partial}{\partial z}\right),
\end{equation}
where $z = x + i y$ is the complex plane coordinate. Introducing the cyclotron frequency $w_c = \frac{e B}{m c}$, we write the Hamiltonian in the form

\begin{equation}
\mathcal{H} = \hbar w_c A^\dagger_e A_e.
\end{equation}
It should be noted that the Hamiltonian has this form for a negatively charged particle. For a positively charged particle, the difference is with respect to the substitution of $A_e$ for $B_e$. This can be easily verified without the need to perform calculations: the sign reversal of the particle is equivalent to the sign reversal of the magnetic field. In turn, the rotation of the magnetic-field axis implies the reversal of all rotations, i.e., the conjugation of complex coordinate $z$. It only remains to be noted that operators $A^\dagger_e$ and $B^\dagger_e$ differ exactly by the complex conjugation. Therefore, we changed the definition of the creation and annihilation operators for negatively charged particles $A^\dagger_h = B^\dagger_e$ and $B^\dagger_h = A^\dagger_e$.
In the situation of interest, there are two Fermi holes and one electron. Hence, the three-particle Hamiltonian takes the form
\begin{equation}
\mathcal{H} = \sum_{i = 1,2}H_{0h}(\mathbf{r_i}) + H_{0e}(\mathbf{r_e}) + \sum_{i = 1,2}U_{eh}(\mathbf{r_i} - \mathbf{r_e}) + U_{hh}(\mathbf{r_1} - \mathbf{r_2})
\end{equation}

The creation operator associated with the exact symmetry of the Hamiltonian, namely the magnetic translation, may be written as (\ref{creation_operator}).
\begin{widetext}
\begin{equation}
\label{creation_operator}
\widetilde{B}^\dagger = B^\dagger_h(\mathbf{r_1}) + B^\dagger_h(\mathbf{r_2}) - B_e(\mathbf{r_e}) = \frac{1}{\sqrt{2}} \left[\frac{z_1 + z_2 - z_e}{2 l_B} - 2 l_B \left(\frac{\partial}{\partial z_1^{*}} + \frac{\partial}{\partial z_2^{*}} + \frac{\partial}{\partial z_e^{*}}\right) \right]
\end{equation}
\end{widetext}
This operator commutes with the free part of the Hamiltonian as it is dependent only on operators $A$, and it also commutes with all paired interactions that are dependent only on the particle coordinate difference. For computational convenience, let us substitute variables $\{r_{1}, r_{2}, r_{e}\} \rightarrow \{r, R, r_e\}$ , where $r = (r_1 - r_2)$.
Then $\widetilde{B}^\dagger = \sqrt{2}B^\dagger_h(\mathbf{R}) - B^\dagger_e(\mathbf{r_e})$. To directly introduce this exact Hamiltonian symmetry in our basis, we performed a Bogoliubov transformation with a unitary operator
$\widetilde{S} = \exp(\Theta\widetilde{L})$, where $\widetilde{L} = B^\dagger_h(\mathbf{R})B^\dagger_e(\mathbf{r_e}) - B_h(\mathbf{R})B_e(\mathbf{r_e})$ and $\cosh(\Theta) = \sqrt{2}$. After the transformation, the creation and annihilation operators take the form
\begin{equation}
\widetilde{B}^\dagger_h(\mathbf{R}) = \sqrt{2}B^\dagger_h(\mathbf{R}) - B^\dagger_e(\mathbf{r_e})
\end{equation}
\begin{equation}
\widetilde{B}^\dagger_e(\mathbf{r_e}) = \sqrt{2}B^\dagger_e((\mathbf{r_e})) - B^\dagger_h(\mathbf{R})
\end{equation}
\begin{widetext}
\begin{equation}
\label{new_basis}
\frac{A_h^{\dagger}(\mathbf{r})^{n_r}A_h^{\dagger}(\mathbf{R})^{n_R}A_e^{\dagger}(\mathbf{r}_e)^{n_e}\widetilde{B}_h^{\dagger}(\mathbf{R})^{k}\widetilde{B}_e^{\dagger}(\mathbf{r}_e)^{l}B_h^{\dagger}(\mathbf{r})^{m}}{(n_r! n_R! n_e! k! l! m!)^{1/2}}|\widetilde{0}\rangle = | n_r n_R n_e; \widetilde{k l m}\rangle
\end{equation}
\begin{equation}
\label{matrix_elements}
\langle \widetilde{m_2 l_2 0}; 1 0 0 |H_{int}| 0 0 1; \widetilde{0 l_1 m_1}\rangle \equiv \langle m_2 l_2|H_{int}|l_1 m_1\rangle = \langle \overline{m_2 l_2 0; 1 0 0} |\overline{S}^\dagger H_{int}\overline{S}|\overline{ 0 0 1; 0 l_1 m_1}\rangle
\end{equation}
\begin{equation}
\label{interaction_hamiltonian}
H_{int} = H_{hh} + H_{eh} = \frac{e^2}{\sqrt{2} \epsilon r} -\frac{\sqrt{2}e^2}{\epsilon |\rho_2 - r|}-\frac{\sqrt{2}e^2}{\epsilon |\rho_1 - r|}.
\end{equation}
\end{widetext}
Hence, we obtain a new basis (\ref{new_basis}), where $|\widetilde{0}\rangle = \widetilde{S}^\dagger|0\rangle$. In the new basis, allowance is made for the exact symmetry of the problem: the Hamiltonian is degenerate in quantum number $k$, and allowance is also made for three approximate symmetries in a strong magnetic field; that is, to the first order of perturbation theory, the Hamiltonian is diagonal in terms of the number of original Landau levels for particles $(n_r, n_R, n_e)$. The numerical diagonalization of the Hamiltonian by the two remaining quantum numbers is no longer of any particular complexity. However, it is necessary to calculate the matrix elements of the interacting part of the Hamiltonian, which is not easily achieved by direct calculations. To simplify the calculation, we performed another Bogoliubov transformation, which will enable the calculation of Coulomb interaction matrix elements in a completely analytical form.
The second Bogoliubov transformation is similar to the first, only for operators
$A^\dagger$:  $\overline{S} = \exp(\Theta\overline{L})$, $\overline{L} = A^\dagger_h(\mathbf{R})A^\dagger_e(\mathbf{r_e}) - A_h(\mathbf{R})A_e(\mathbf{r_e})$. The consistent application of these transformations is equivalent to a change of coordinates $\{r, R, r_e\} \rightarrow \{ r, \rho_1, \rho_2\}$, where $\rho_1 = \sqrt{2}R - r_h$ and $\rho_2 = \sqrt{2}r_h - R$. Considering the trion structure, we are interested in $n_r = 0$, $n_R = 0$, $n_e = 1$. The matrix elements for the interaction part of the Hamiltonian take the form (\ref{matrix_elements}) with interaction Hamiltonian (\ref{interaction_hamiltonian}). We obtain final expressions for matrix elements (\ref{final_expression_1}, \ref{final_expression_2}, \ref{eq:Ueh}).
\begin{widetext}
\begin{equation}
\label{final_expression_1}
\langle l_1 m_1| U_{ee}(\sqrt{2}r) |l_2 m_2 \rangle = \delta_{m_1, m_2} \delta_{l_1, l_2} \frac{1}{l_B^2m!} \int\limits_{0}^{\infty}{ r dr \left( \frac{r}{\sqrt{2}l_B} \right)^{2m} \exp\left(-\frac{r^2}{2 l_B^2}\right) U_{ee}(\sqrt{2}r)}
\end{equation}
\begin{equation}
\label{final_expression_2}
\langle m_2 l_2|H_{eh}|l_1 m_1\rangle = 2 \sqrt{2} \bar U_{0 m_2 0 l_2}^{0 m_1 0 l_1} =
\delta_{l_1 - m_1, l_2 - m_2}
\bar U_{\min(m_1, m_2), \min(l_1, l_2)}(|m_1 - m_2|)
\end{equation}
\begin{equation}\label{eq:Ueh}
\bar U_{mn}(s) = {E_0 \sqrt{\frac{2}{\pi}}
	\sqrt{\frac{m! n!}{(m + s)!(n + s)s!}}
	\int\limits_{0}^{\infty}{d q}
	f(q) 2(1-2x) e^{-3x} x^{s} L^s_m(x) L^s_n(x)}
\end{equation}
\end{widetext}
In the above formulas, $E_0$ is the single-magnetoexciton binding energy in a $\delta$-functional quantum well, $f(q)$ is the geometric form factor related to the nonlocality of the electron and Fermi-hole wave functions in the QW growth direction, $L_m^s(x)$ is the generalized Laguerre polynomial, and $x = q^2 l_B^2 / 2$. We chose a basis size for Hamiltonian diagonalization $60^{2} \times 60^{2}$. With a further increase in the number of basis functions the computational accuracy increases slowly owing to the rapid oscillations of the integrand in (\ref{eq:Ueh}). 

When calculating the energy spectrum, consideration must be given to obtain the correct parity of trion wave functions for singlet, $S_h = 0$, and triplet, $S_h = 1$, two-hole states. It is also important that the total orbital momentum of the trion, $M_z = n_r + n_R - n_e - k + l - m$, is kept, which enables us to seek bound trion states over sectors with fixed values of $M_z$. The dipole-allowed transitions for the trion inside the conduction band are those with $\Delta M_z = \pm 1$. However, optical transitions of the photoexcited hole from the valence band to the conduction band require the conservation of the orbital momentum $\Delta M_z = 0$. It is also assumed that during optical transition, the magnetic trion translation vector and spins of magnetoexciton electron and hole are kept constant. It appears that even considering the chosen assumptions, the spectrum of three-particle states exhibits multiple bound states that correspond to the same $M_z$ (Figure~\ref{fig:fig2}). The energy difference between the nearest bound states with equal principal quantum numbers of trions falls in the energy range separating basic dipole-allowed trion transitions and new PL lines (Figure~\ref{fig:fig3}). Thus, it is established that the new lines in the PL spectra of the 2D electron system in the presence of a dense ensemble of nonequilibrium long-lived triplet magnetoexcitons are associated with the complex intrinsic motion of a trion. The calculations show that the maximum number of lines that can be seen in the PL spectra does not exceed four. To date, the reason for the violation of internal trion symmetry during the transformation of a photoexcited valence-band hole to a Fermi hole is unknown, and it is therefore difficult to predict the lines that should be either more or less active in the PL spectra. Accordingly, it is not known which of the feasible transitions with $\Delta M_z = 0$ are associated with the observed spectral lines. This may be determined by further clarifying the theory of three-particle states in a quantizing magnetic field in the presence of a random potential.

\section{Plasmarons}

The investigation of plasmarons is a far more complex problem than the previous one, which is due to the absence of an available theoretical description of three-particle complexes interacting with magnetoplasma oscillations in a 2D electron system \cite{Dickmann2005}. Therefore, in this section, we focus on experimental results that explain peculiarities of plasmaron physics. It will be shown that there exist not just one plasmaron with fixed energy, but an entire band of plasmaron excitations. The difference between plasmarons is related to the initial generalized momentum of the triplet magnetoexciton forming the plasmaron.

\begin{figure}[h!]
	\includegraphics{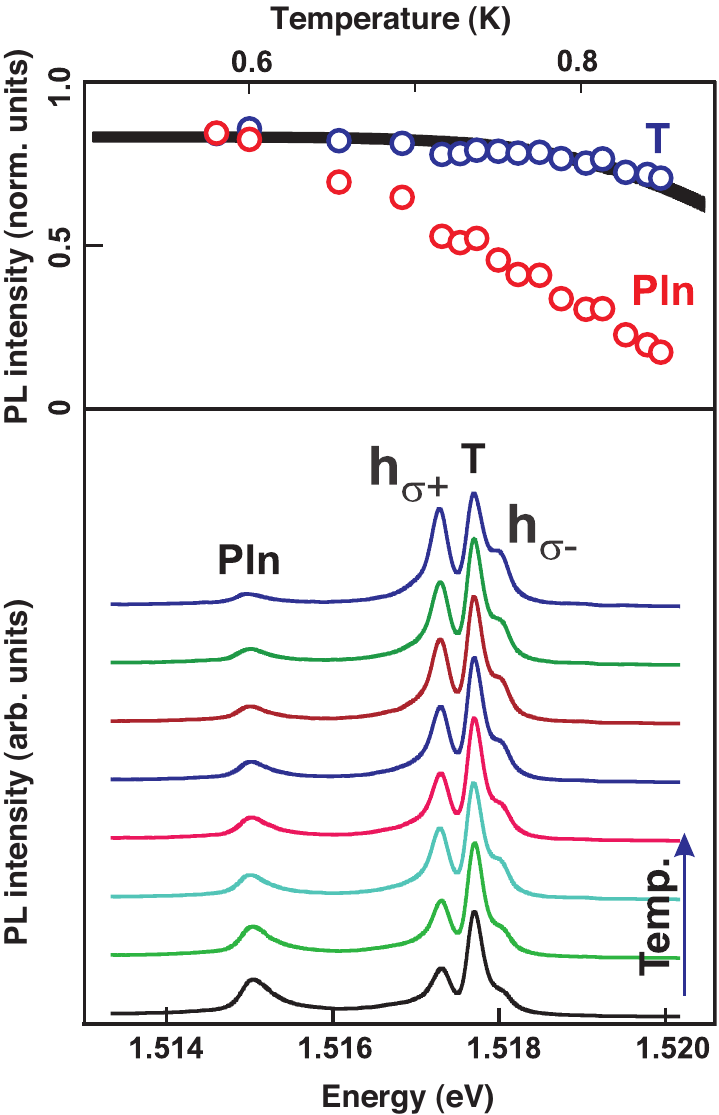}
	\caption{\label{fig:fig4}
		The temperature dependence of PL spectra (bottom) in the sample with concentration $n_{e}=2\cdot 10^{11}$ cm$^{-2}$. The top graph shows ratios of integral intensities of three-particle and single-particle transitions. The solid line represents the reduction of the magnetoexciton density as a function of temperature obtained by the best approximation of the experimental data of [\cite{Kulik2016}].}
\end{figure}

As distinct from trions, plasmarons are more fragile formations with binding energy lower than that of the trion, as evidenced from the temperature dependence of plasmaron and trion line intensities and the intensities of corresponding single-particle lines (Figure~\ref{fig:fig4}). The intensities of both trion and plasmaron lines decrease with increasing temperature. The trion-line intensity follows the temperature dependence of the triplet magnetoexciton number that was obtained for the sample under study \cite{Kulik2016}. Thus, the temperature dependence of the trion line intensity does not reflect the binding energy of the trion itself, but demonstrates the reduction of the magnetoexciton density only. This confirms the calculations of the previous section, which show that the trion binding energy (1.5 meV) is significantly higher than the activation gap for the ``ionization'' of triplet magnetoexcitons.

As for the plasmaron line intensity, it falls more rapidly with increasing temperature, and at 1 K, there are practically no plasmarons left in the electron system. Assuming that the plasmaron binding energy is substantially lower than the magnetoplasma oscillation energy, this energy can be omitted for the plasmaron description, and the third charge (Fermi hole) is considered almost free. Its role is significant for magnetoexciton-magnetoplasmon momentum transfer (for the generalized momentum conservation), but after the formation of a magnetoplasmon, the effect of the Fermi hole motion on magnetoplasma oscillations can be omitted. Accordingly, the number of different plasmarons will be the same as the number of possible states of triplet magnetoexcitons with different momenta, whereas the maximum number of plasmarons should possess energy values that correspond to that of magnetoplasma oscillations at the generalized momentum matching the minimum of the dispersion for triplet magnetoexcitons.

\begin{figure}[h!]
	\includegraphics{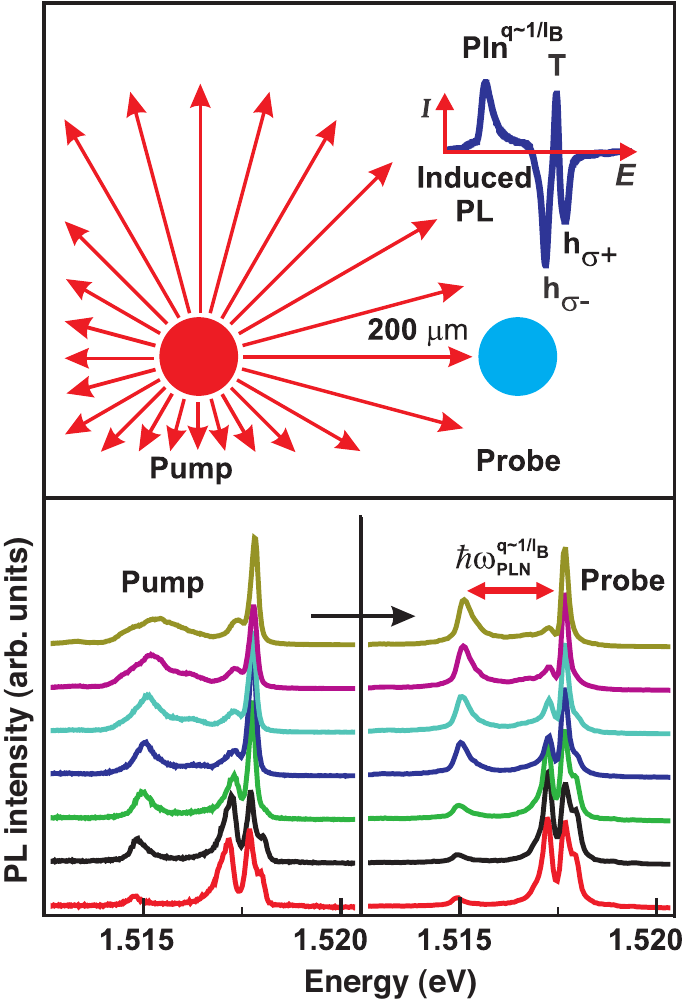}
	\caption{\label{fig:fig5}
		Effect of photoexcitation power $P$ on PL spectra ($n_{e}=2\cdot 10^{11}$ cm$^{-2}$). (top) Scheme of experiment. (bottom) Series of PL spectra in the pump spot (left) and in the probe spot (right). The graph in top panel shows the difference of the PL spectra registered in the probe spot at $P=$ 200 $\mu$W and $P=$ 0. The single-particle transition lines are reduced dramatically, while the lines of trion and plasmaron (with the energy corresponding to the minimum in the dispersion curve of triplet magnetoexciton, $q\sim 1/l_B$) are increased. However, in the pump spot the spectral share increases of plasmarons with all other energies due to the escape from the excitation region of magnetoexcitons with $q\sim 1/l_B$.}
\end{figure}

\begin{figure}[h!]
	\includegraphics{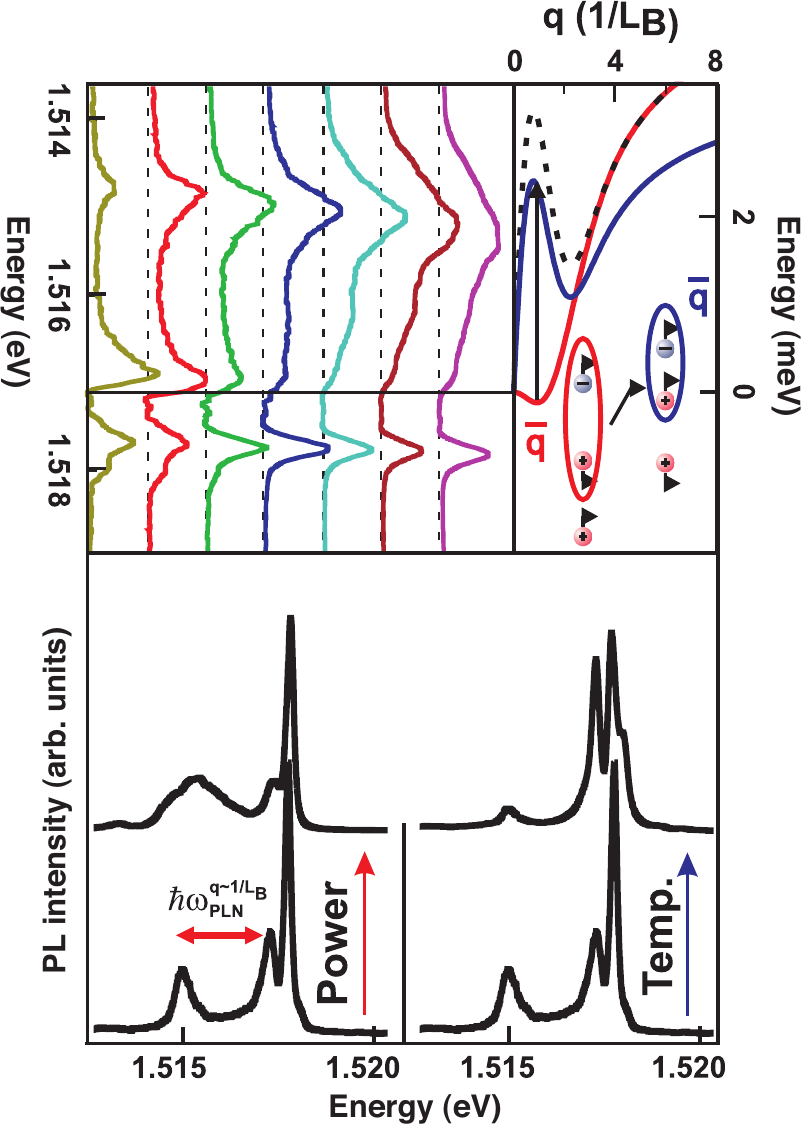}
	\caption{\label{fig:fig6}
		Assumed momentum distribution (top left) of magnetoplasmons (plasmarons) as a function of the pump power obtained from the pump-spot PL spectra in $\sigma^{-}$ polarization after subtraction of single-particle peak. Dispersion dependences (top right) of triplet magnetoexciton and magnetoplasmon of the sample in question, calculated in the first order of perturbation theory considering the confining potential profile in the growth direction for the QW (red and black lines). The blue line shows the expected dispersion dependence of the magnetoplasmon accounting for the Coulomb influence of the extra Fermi hole on the plasmaron energy discovered in [\cite{Zhuravlev2016}]. Effect of both excitation power and temperature on the PL spectra in the pump spot (bottom).}
\end{figure}

To verify the above assumption, we performed an experiment on triplet magnetoexciton transport at macroscopic distances. A dense ensemble of triplet magnetoexcitons was created in a pump spot. The transport of magnetoexcitons from the pump spot was monitored by the measurement of PL spectra in a probe spot spaced 200-$\mu$m apart from the pump spot. The power of photoexcitation (background number of triplet magnetoexcitons in quasi-equilibrium conditions) in the probe spot remained constant, whereas the photoexcitation power in the pump spot changed by two orders of magnitude. At a weak excitation level in the pump spot (less than 6 $\mu$W), the form of the PL spectrum in the probe spot does not change, which suggests that the triplet magnetoexciton ensemble in the pump spot does not affect the electron states in the probe spot. As the pump power exceeds some threshold value that is generally related to the formation of ``magnetofermionic condensate'' in the triplet magnetoexciton ensemble \cite{Kulik2016}, the probe spot exhibits a change in the PL signal (Figure~\ref{fig:fig5}). This means that magnetoexcitons start to arrive at the probe spot from a distance of 200 $\mu$m. Recently, we have reported on visualization of spin-triplet magnetoexciton spreading over macroscopic distances using technique of photo-induced resonant reflection \cite{Kulik2018}. We suppose that the maximal observed magnetoexciton propagation distance ($\geq$ 200 $\mu$m) is limited then and now by the optical aperture of our experimental setup. Here, the optical transitions with participation of three-particle complexes play role of ``signal labels'' for appearance of pump-spot triplet magnetoexcitons in the probe spot. Considering the difference between the PL spectra in the probe spot in the presence and absence of excitation in the pump spot, we can obtain the pump-induced PL signal. The PL signal induced by the spatially separated pump consists of a positive signal from three-particle complexes, trions and plasmarons, and a negative signal from single-particle transitions. In particular, a large enhancement in the signal of plasmarons with the energy corresponding to the minimum in the dispersion dependence of triplet magnetoexcitons, $q\sim 1/l_B$, is observed (Figure~\ref{fig:fig5}).

Since the integrated luminosity remains unchanged, the escape from the pump spot of magnetoexcitons with minimum possible energy (which corresponds to magnetoplasmon (plasmaron) energies with the generalized momentum $\sim 1/l_B$) should increase the luminosity of plasmarons with other energies. This is verified by measuring the PL spectra in the pump spot. To reduce the contribution of trion lines, which are out of use at the investigation of plasmarons, the spectra were measured in $\sigma^{-}$  polarization with a small portion of $\sigma^{+}$ polarization in order to control the trion line energy. Besides, the PL line of single-particle transitions in $\sigma^{-}$  polarization was approximated by a Gaussian curve and subtracted from the PL spectra. This procedure enables the visualization of magnetoplasmon, and correspondingly, the plasmaron distribution in the allowed states defined by the triplet magnetoexciton dispersion (Figure~\ref{fig:fig6}). Under the conditions of the escape of triplet magnetoexcitons with minimal energy from the pump spot, the energy of observed plasmaron excitations is shifted to the violet side and forms a practically continuous band right up to the line of single-particle optical transitions (Figure~\ref{fig:fig6}). Moreover, additional high-energy plasma oscillations appear, which are likely to be associated with magnetoplasma modes filling the tail of dispersion curves with $ql_B > 2$. Note that the change of the magnetoexciton distribution in the pump spot is not due to the overheating of the electron system under the pump, which is directly demonstrated by the temperature dependence of the intensity of the plasmaron PL at a fixed pump power (Figure~\ref{fig:fig6}).

\section{Conclusion}

In conclusion, we present the key results of this study. Further studies have been made to determine the properties of three-particle complexes discovered in [\cite{Zhuravlev2016}]. It is experimentally shown that the PL spectrum of the 2D system in the quantum Hall insulator reveals new lines linked to the formation of a nonequilibrium triplet magnetoexciton ensemble. The spectrum of bound three-particle states formed by two Fermi holes and a photoexcited electron was theoretically calculated. It was found that at the same orbital quantum number of a trion, there exist not one but several excited bound states distinguished by additional quantum numbers of intrinsic motion of the particles that form the trion. This enables optical transitions from the valence band to the conduction band with an unchanged orbital quantum number, but different energies. The energies of feasible optical transitions fall within the energy range of new spectral lines, which indicates their origins: optical transitions with unchanged principal quantum numbers (vector of magnetic translations, orbital quantum number, electron and holes spins) and the violation of intrinsic quantum numbers of trions related to the motion of an electron and holes around the center of mass.

Experiments were performed on the spatial transfer of the triplet magnetoexciton density at macroscopic distances, which showed that three-particle complexes interacting with magnetoplasma oscillations in the electron system form a continuous band of collective excitations. The energy of the latter is determined by the initial generalized momentum of triplet magnetoexcitons that form the plasmaron. The generalized momentum transfer from a two-particle to three-particle complex in the magnetic field can be explained by the low binding energy for the plasmaron when compared to the magnetoplasma oscillation energy and the participation of the extra Fermi hole as a mediator of the transfer process. Such a process would be impossible without the extra Fermi hole because the spins of the electron and Fermi hole in the spin-triplet magnetoexciton are different. Therefore, the formation of plasmaron requires the delocalization of the in-magnetoexciton Fermi hole and the incorporation of another Fermi hole with a required spin into a magnetoplasmon with the generalized momentum of the original magnetoexciton.

\begin{acknowledgments}
This research was supported by the Russian Science Foundation Grant No. 16-12-10075. I. V. K thanks Russian Fund for Fundamental Research for computational facilities. We thank D. Yakovlev and S. Dickman for valuable discussions.
\end{acknowledgments}

\end{document}